\documentclass{easychair}

\usepackage{url}
\usepackage{amssymb}
\usepackage{amsmath}
\usepackage{alltt}
\usepackage{mathtools}
\usepackage{dsfont}
\usepackage[T1]{fontenc}
\usepackage{fancybox} 
\usepackage{titlesec}

\mathcode`:="003A  
\mathcode`;="003B  
\mathcode`?="003F  
\mathcode`|="026A  
\mathcode`<="4268  
\mathcode`>="5269  

\mathchardef\ls="213C    
\mathchardef\gr="213E    

\newenvironment{todo}{\bigskip\hrule\medskip\noindent}{\medskip\hrule\bigskip}

\begin{document}
\bibliographystyle{abbrv}

\titlerunning{Coffman deadlocks in SCOOP}
\title{Coffman deadlocks in SCOOP\footnote{The research leading to these results has received funding from the
European Research Council under the European Union's Seventh Framework
Programme (FP7/2007-2013) / ERC Grant agreement no. 291389.}}

\author{Georgiana Caltais\inst{1} \and Bertrand Meyer\inst{1,2,3}}
\authorrunning{G. Caltais \and B. Meyer }

\institute{
Department of Computer Science, ETH Z\"urich, Switzerland \and
Eiffel Software, Santa Barbara \and
NRU ITMO, Saint Petersburg
}

\maketitle 

{\begin{abstract}
In this paper we address the deadlock detection problem in the context of SCOOP -- an OO-programming model for concurrency, recently formalized in Maude. We present the integration of a deadlock detection mechanism on top of the aforementioned formalization and analyze how an abstract semantics of SCOOP based on a notion of ``may alias expressions'' can contribute to improving the deadlock detection procedure. 
\end{abstract}
}

\vspace{-10pt}
\paragraph{Introduction.}
{

In this paper we are targeting SCOOP~\cite{DBLP:conf/acsd/MorandiSNM13} -- a concurrency model recently provided with a formalization based on Rewriting Logic (RL)~\cite{DBLP:conf/fct/MeseguerR11}, which is ``executable'' and straightforwardly implementable in the programming language Maude. 
\emph{Our aim} is to develop a (Coffman) deadlock~\cite{Coffman:1971:SD:356586.356588} detection mechanism for SCOOP applications. Intuitively, such deadlocks occur whenever two or more executing threads are each waiting for the other to finish.

\emph{Our contribution.} We present the integration of a deadlock detection mechanism on top of the formalization in~\cite{DBLP:conf/acsd/MorandiSNM13}. We also briefly analyze how a simplified, abstract semantics of SCOOP based on a notion of ``may alias'' expressions~\cite{Meyer-aliasing-13,Caltais-aliasing} can be exploited in order to improve the deadlock detection procedure.
}

The literature on using static analysis~\cite{Landi:1992:USA:161494.161501} and abstracting techniques for (related) concurrency models is considerable. We refer, for instance, to the recent work in~\cite{DBLP:conf/ifm/GiachinoGLLW13} that introduces a framework for detecting deadlocks by identifying circular dependencies in the (finite state) model of so-called contracts that abstract methods in an {OO}-language. The integration of a deadlock analyzer in SCOOP on top of Maude is an orthogonal approach that belongs to a more ambitious goal, namely the construction of a RL-based toolbox for SCOOP programs including a \emph{may alias} analyzer, as thoroughly presented in~\cite{Caltais-aliasing} and a type checker.

\vspace{-10pt}
\paragraph{Deadlock detection in SCOOP.}
{

The key idea of SCOOP is to associate to each object a processor, or \emph{handler} (that can be a CPU, or it can also be implemented in software, as a process or thread). In SCOOP terminology, objects that can run on different processors are \emph{separate} from each other. Assume a processor $p$ that performs a call $o.f(a_1, a_2, \ldots)$ on an object $o$. If $o$ is declared as ``separate'', then $p$ sends a request for executing $f(a_1, a_2, \ldots)$ to $q$ -- the handler of $o$ (note that $p$ and $q$ can coincide). Meanwhile, $p$ can continue. Moreover, assume that $a_1, a_2, \ldots$ are of ``separate'' types. In the SCOOP semantics, the application of the call  $f(\ldots)$ will \emph{wait} until it has been able to \emph{lock} all the separate objects associated to $a_1, a_2, \ldots$. This mechanism guarantees exclusive access to these objects. Processors communicate via \emph{channels}.

In the context of SCOOP,  the \emph{deadlocking problem} reduces to identifying whether a set of processors reserve each other circularly. This situation might occur, for instance, in a Dining Philosophers scenario, where both philosophers and forks are objects residing on their own processors.
Given a processor $p$, by $W(p)$ we denote the set of processors $p$ \emph{waits} to release the resources $p$ needs for its asynchronous execution. Orthogonally, by $H(p)$ we represent the set of resources (more precisely, resource handlers that) $p$ already acquired. We say that a deadlock exists if for some set $D$ of processors:
$
(\forall p \in D).(\exists p' \in D).(p \not = p').W(p) \cap H(p') = \emptyset~(\clubsuit).
$

The semantics of SCOOP in~\cite{DBLP:conf/acsd/MorandiSNM13} is defined over tuples of shape   
$
\langle p_1 \,::\, St_{1} \mid \ldots \mid p_n \,::\, St_{n}, \sigma \rangle
$
where, $p_i$ denotes a processor (for $i \in \{1, \ldots, n\}$), $St_i$ is the call stack of $p_i$ and $\sigma$ is the {\it state} of the system. States hold information about the {\it heap} (which is a mapping of references to objects) and the {\it store} (which includes formal arguments, local variables, {\it etc}.).
Integrating the deadlock definition in~$(\clubsuit)$ on top of the Maude formalization in~\cite{DBLP:conf/acsd/MorandiSNM13} is almost straightforward. Given a processor $p'$ as in~$(\clubsuit)$, the set $H(p')$ corresponds, based on~\cite{DBLP:conf/acsd/MorandiSNM13}, to $\sigma.rq\_locks(p')$.
Whenever the top of the instruction stack of a processor $p$ is of shape $lock(\{q_i,\ldots,q_n\})$, we say that the wait set $W(p)$ is the set of processors $\{q_1, \ldots, q_n\}$.
Hence, assuming a predefined system configuration $\langle deadlock \rangle$, the SCOOP transition rule in Maude corresponding to~$(\clubsuit)$ can be written as:
\vspace{-10pt}
\begin{equation}
\label{eq:Maude-deadlock}
\dfrac{
\begin{array}{c}
(\exists D \subseteq \sigma.procs).(\forall p \in D). (\exists p' \in D).(p\not = p').\\
(aqs\,:=\ldots \mid p\,::\,lock(\{q_i,\ldots\});St \mid \ldots)\,\,\land\,\,
(\sigma.rq\_locks(p').has(q_i))
\end{array}
}
{
\langle aqs, \sigma\rangle \rightarrow \langle deadlock \rangle
} 
\end{equation}
It is intuitive to guess that $\sigma.procs$ in~(\ref{eq:Maude-deadlock}) returns the set of processors in the system, whereas $aqs$ stands for the list of these processors and their instruction stacks (separated by the associative \& commutative operator ``$\mid$''~). We use ``$\ldots$'' to represent an arbitrary sequence of processors and processor stacks.
}

\vspace{-10pt}
\paragraph{Discussion.}
{
We implemented~(\ref{eq:Maude-deadlock}) and tested the deadlock detection mechanism on top of the formalization in~\cite{DBLP:conf/acsd/MorandiSNM13} for the Dining Philosophers problem. A case study considering two philosophers can be run by downloading the SCOOP formalization at:\\
{\footnotesize\verb+https://dl.dropboxusercontent.com/u/1356725/SCOOP-NWPT-14.zip+}, 
and executing the command\\
{\footnotesize\verb+> maude SCOOP.maude ..\examples\dining-philosophers-example.maude+}.
In our example, 
the philosophers {\footnotesize\verb+p1+} and {\footnotesize\verb+p2+} can reach a (Coffman) deadlock ({\footnotesize\verb+go_wrong(p1, p2)+}) if they adopt a wrong eating strategy ({\footnotesize\verb+pi.eat_wrong+}, with {\footnotesize\verb+i+} $\in \{1, 2\}$). This might happen whenever a philosopher proceeds by picking up the forks {\footnotesize\verb+f1+} and {\footnotesize\verb+f2+} on the table in turn ({\footnotesize\verb+pick_in_turn(fi)+}, with {\footnotesize\verb+i+} $\in \{1, 2\}$) instead of picking them at the same time.
It might be the case that, for instance, {\footnotesize\verb+p1+} picks up {\footnotesize\verb+f1+}, whereas {\footnotesize\verb+p2+} immediately picks up {\footnotesize\verb+f2+}. Thus, each of the two philosophers is holding a fork the other philosopher is waiting for.

As can be seen from the code in {\footnotesize \verb+dining-philosophers-example.maude+}, in order to implement our applications in Maude, we use intermediate representations.
For a brief example, consider the class implementing the \emph{philosopher} concept, given below:
\vspace{-25pt}
\paragraph{}{
\footnotesize{
\begin{verbatim}
(class 'PHILOSOPHER 
    create { 'make } (
        attribute {'ANY} 'left : [!,T,'FORK] ; attribute {'ANY} 'right : [!,T,'FORK] ;
        
        procedure { 'ANY } 'make ( 'fl : [!,T,'FORK] ; 'fr : [!,T,'FORK] ; ) 
            do  ( assign ('left, 'fl) ; assign ('right, 'fr) ; )
            [...]
        end ;
[...] end)
\end{verbatim}
}
}
\vspace{-5pt}
\noindent
it declares two forks -- {\footnotesize\verb+'left+} and {\footnotesize\verb+'right+} of type {\footnotesize\verb+[!, T, 'FORK]+}, that can be handled by any processor ({\footnotesize\verb+T+}) and that cannot be {\footnotesize\verb+Void+} ({\footnotesize\verb+!+}). The corresponding constructor {\footnotesize\verb+'make('fl, 'fr)+} initializes the philosopher's forks accordingly.

It is worth pointing out that in the aforementioned example we use a predefined strategy~\cite{Marti-OlietMV05} that guides the rewriting of the Maude rules formalizing SCOOP towards a $\langle deadlock \rangle$ system configuration.
Nevertheless, such an approach requires lots of ingeniousness and, moreover, is not automated. Given the size of the current SCOOP formalization, running the Maude model checker is, unfortunately, not an option. We anticipate a ``way out'' of the state explosion issue by exploiting the expression-based alias calculus in~\cite{Meyer-aliasing-13} in order to provide a simplified, abstract semantics of SCOOP. 
In short, the calculus in~\cite{Meyer-aliasing-13} identifies whether two expressions in a program \emph{may} reference to the same object. Consider, for intuition, the code {\footnotesize \verb+x := y; loop x := x.next end+} that assignes a linked list. The corresponding execution causes {\footnotesize \verb+x+} to become aliased to {\footnotesize \verb+y.next.next. ...+}, with a possibly infinite number of occurrences of the field {\footnotesize \verb+next+}. The set of associated ``may alias'' expressions identified by the calculus in~\cite{Meyer-aliasing-13} can be equivalently written as $\{ [$ {\footnotesize \verb+x+}, {\footnotesize \verb+y.next+} $\!\!^k]\mid k \geq 0 \}$.

The idea behind using an alias-based abstract semantics of SCOOP stems from the fact that SCOOP processors are known from object references, which may be aliased. Therefore, the SCOOP semantics could be simplified by retaining within the corresponding transition rules only the information relevant for aliasing. Consider, for instance, the assignment instruction formally specified as:
\vspace{-5pt}
\[
\dfrac{\textnormal{a is fresh}}{
\langle p \,::\, t\,:=s;\, St,\, \sigma \rangle \rightarrow
\langle p\,::\, \textnormal{eval}(a, s);\, \textnormal{wait}(a);\, \textnormal{write}(t, a.data);\,St,\, \sigma\rangle
}~.
\]
Intuitively, ``eval$(a, s)$'' evaluates $s$ and puts the result on channel $a$, ``wait$(a)$'' enables processor $p$ to use the evaluation result and ``write$(t, a.data)$'' sets the value of $t$ to $a.data$. The abstract transition rule
omits the evaluation of the right-hand side of the assignment $t\,:=s$ and the associated message passing between channels, and updates the aliasing information in the newly added component \emph{alias\_} (consisting of a set of alias expressions) according to the calculus in~\cite{Meyer-aliasing-13}:
\vspace{-5pt}
\[
{
\dfrac{\cdot}{
\langle p \,::\, t\,:=s;\, St,\, \sigma, alias_{old} \rangle \rightarrow
\langle p\,::\, \,St,\, \sigma, alias_{new} \rangle
}}~.
\]
\noindent
Then, the rule~(\ref{eq:Maude-deadlock}) identifying deadlocks can be naturally redefined to range over the expressions aliased with the processors $p, p'$ and $q_i$, respectively. Nevertheless, observe that this approach is prone to introducing ``false positives'' w.r.t. the expressions that would actually become aliased at runtime; this is due to the over-approximating nature of the alias calculus in~\cite{Meyer-aliasing-13} that ignores conditions 
in conditionals and loops.
Furthermore, the abstract setting enables the simplification of the SCOOP semantics by completely eliminating the rules formalizing the exception handling mechanism, for instance.
We plan to closely investigate and implement this abstraction mechanism in Maude.
For a survey on similar ``abstracting'' procedures we refer to the work in~\cite{DBLP:conf/fct/MeseguerR11}.
}

\vspace{-10pt}

\begingroup
\titleformat*{\section}{ \fontsize{10}{12.5}\selectfont\bf}


\newpage
\appendix
\section{Dining Philosophers in SCOOP}
\label{app-example}
In what follows, we provide the relevant parts of the intermediate class-based representation in {\footnotesize\verb+dining-philosophers-example.maude+}, together with parts of the Maude output corresponding to the strategy-based execution of the example:
\vspace{-20pt}
\paragraph{}{
\footnotesize
\begin{verbatim}
srew

(( import default

(class 'APPLICATION 
    create 
        { 'make }
    (
        attribute {'ANY} 'meal : [!,T,'MEAL] ;
        
        procedure { 'ANY } 'make (nil) 
            require True 
            local ( nil )
            do
                (
                create ('meal . 'make(nil)) ;
                command('Current . 'execute_wrong('meal ;)) ;
                )
            ensure True
            rescue  nil
        end ;
        
        procedure { 'ANY } 'execute_wrong ( 'm : [!,T,'MEAL] ;  ) 
            require True 
            local ( nil )
            do
                (
                command('m . 'do_wrong(nil)) ;
                )
            ensure True
            rescue  nil
        end ;
    )
    
    invariant True 
end) ;

(class 'MEAL 
    create 
        { 'make }
    (
        attribute {'ANY} 'p1 : [!,T,'PHILOSOPHER] ;
        attribute {'ANY} 'p2 : [!,T,'PHILOSOPHER] ;
        attribute {'ANY} 'f1 : [!,T,'FORK] ;
        attribute {'ANY} 'f2 : [!,T,'FORK] ;
        
        procedure { 'ANY } 'make (nil) 
            require True 
            local ( nil )
            do
                (
                create ('f1 . 'make(nil)) ;  create ('f2 . 'make(nil)) ;
                create ('p1 . 'make('f1 ; 'f2 ;)) ;  create ('p2 . 'make('f2 ; 'f1 ;)) ;
                )
            ensure True
            rescue  nil
        end ;
        
        procedure { 'ANY } 'do_wrong (nil) 
            require True 
            local ( nil )
            do
                (
                command ('Current . 'go_wrong('p1 ; 'p2 ;)) ;
                )
            ensure True
            rescue  nil
        end ;
        
        procedure { 'ANY } 'go_wrong ('pa : [!,T,'PHILOSOPHER] ; 'pb : [!,T,'PHILOSOPHER] ;) 
            require True 
            local ( nil )
            do
                (
                command ('pa . 'eat_wrong(nil)) ;
                command ('pb . 'eat_wrong(nil)) ;
                )
            ensure True
            rescue  nil
        end ;
    )
    
    invariant True 
end) ;

(class 'PHILOSOPHER 
    create 
        { 'make }
    (
        attribute {'ANY} 'left : [!,T,'FORK] ;
        attribute {'ANY} 'right : [!,T,'FORK] ;
        
        procedure { 'ANY } 'make ( 'fl : [!,T,'FORK] ; 'fr : [!,T,'FORK] ; ) 
            require True 
            local ( nil )
            do
                ( 
                assign ('left, 'fl) ;
                assign ('right, 'fr) ;
                )
            ensure True
            rescue  nil
        end ;
        
        procedure { 'ANY } 'pick_two ('fa : [!,T,'FORK] ; 'fb : [!,T,'FORK] ; ) 
            require True 
            local ( nil )
            do
                (
                command ('fa . 'use(nil)) ;
                command ('fb . 'use(nil)) ;
                )
            ensure True
            rescue  nil
        end ;
        
        procedure { 'ANY } 'eat_wrong (nil) 
            require True 
            local ( nil )
            do
                (
                command ('Current . 'pick_in_turn('left ;)) ;
                )
            ensure True
            rescue  nil
        end ;
        
        procedure { 'ANY } 'pick_in_turn ('f : [!,T,'FORK] ; ) 
            require True 
            local ( nil )
            do
                (
                command ('Current . 'pick_two('f ; 'right ;)) ;
                )
            ensure True
            rescue  nil
        end ;
    )
    
    invariant True 
end) ;

(class 'FORK 
    create 
        { 'make }
    (
        procedure { 'ANY } 'make (nil) 
            require True 
            local ( nil )
            do ( nil )
            ensure True
            rescue  nil
        end ;
        
        procedure { 'ANY } 'use (nil) 
            require True 
            local ( nil )
            do ( nil )
            ensure True
            rescue  nil
        end ;
    )
    
    invariant True 
end) ;

) settings('APPLICATION, 'make, false, deadlock-on)) 
using
init ;  parallelism{lock} ;  [...] ; deadlock-on .
\end{verbatim}
}

The entry point of the program implementing the Dining Philosophers example is the function {\footnotesize\verb+'make+} in the class {\footnotesize\verb+APPLICATION+}. The flag enabling the deadlock analysis is set to ``on''.
This information is specified using the instruction {\footnotesize\verb+settings('APPLICATION, 'make, false, deadlock-on)+}.

A possible scenario that leads to a deadlock when running the above code is as follows. First, we initialize the left and right forks of the philosophers: {\footnotesize\verb+p1+} is assigned {\footnotesize\verb+f1+} and {\footnotesize\verb+f2+}, respectively, whereas {\footnotesize\verb+p2+} is assigned {\footnotesize\verb+f2+} and {\footnotesize\verb+f1+}, respectively. Then, asynchronously, {\footnotesize\verb+p1+} and {\footnotesize\verb+p2+} (of \emph{separate} type {\footnotesize\verb+PHILOSOPHER+}) execute {\footnotesize\verb+eat_wrong+}, which calls {\footnotesize\verb+pick_in_turn(left)+}. In the context of {\footnotesize\verb+p1+}, the actual value of {\footnotesize\verb+left+} is {\footnotesize\verb+f1+}, whereas for {\footnotesize\verb+p2+} it is {\footnotesize\verb+f2+}. Consequently, both resources {\footnotesize\verb+f1+} and {\footnotesize\verb+f2+}, respectively, might be locked ``at the same time'' by {\footnotesize\verb+p1+} and {\footnotesize\verb+p2+}, respectively. Note that {\footnotesize\verb+pick_in_turn+} subsequently calls {\footnotesize\verb+pick_two+} that, intuitively, should enable the philosophers to use both forks. Thus, if {\footnotesize\verb+f1+} and {\footnotesize\verb+f2+}, respectively, are locked by {\footnotesize\verb+p1+} and {\footnotesize\verb+p2+}, respectively, the calls {\footnotesize\verb+pick_two(f2, f1)+} and {\footnotesize\verb+pick_two(f1, f2)+} corresponding to {\footnotesize\verb+p1+} and {\footnotesize\verb+p2+} will (circularly) wait for each other to finish. According to the SCOOP semantics,
{\footnotesize\verb+pick_two(f1, f2)+} is waiting for {\footnotesize\verb+p2+} to release {\footnotesize\verb+f2+}, whereas
{\footnotesize\verb+pick_two(f2, f1)+} is waiting for {\footnotesize\verb+p1+} to release {\footnotesize\verb+f1+}, as the forks are passed to {\footnotesize\verb+pick_two(...)+} as \emph{separate} types. In the context of SCOOP, this corresponds to a Coffman deadlock~\cite{Coffman:1971:SD:356586.356588}. 

We force the execution of the scenario above by applying the command/strategy\\ 
{\footnotesize\verb+srew [...] using init ; parallelism{lock} ; [...] ; deadlock-on+}.
This determines Maude to first trigger the rule {\footnotesize\verb+[init]+} in the SCOOP formalization. This makes all the required initializations of the \emph{bootstrap} processor. Then, one of the processors that managed to \emph{lock} the necessary resources is (``randomly'') enabled to proceed to the asynchronous execution of its instruction stack, according to the strategy  {\footnotesize\verb+parallelism{lock}+} . The last step of the strategy calls the rule  {\footnotesize\verb+[deadlock-on]+} implementing the Coffman deadlock detection as in~(\ref{eq:Maude-deadlock}).
(For a detailed description of SCOOP and its Maude formalization we refer the interested reader to the work in~\cite{DBLP:conf/acsd/MorandiSNM13}.)

We run the example by executing the command:\\
{\footnotesize
\verb+ > maude SCOOP.maude ..\examples\dining-philosophers-example.maude+
}\\
\noindent
The rewriting guided according to the aforementioned strategy leads to one solution identifying a Coffman deadlock.
The relevant parts of the corresponding Maude output are as follows:
\vspace{-20pt}
\paragraph{}{
\footnotesize
\begin{verbatim}
                     \||||||||||||||||||/
                   --- Welcome to Maude ---
                     /||||||||||||||||||\
            Maude 2.6 built: Mar 31 2011 23:36:02
            Copyright 1997-2010 SRI International
                   Wed Sep 17 14:47:47 2014
[...]
==========================================
srewrite in SYSTEM : (import default __create__invariant_end(...) ;
    __create__invariant_end(...) ; __create__invariant_end(...) ;
    __create__invariant_end(...) ;) settings('APPLICATION, 'make, false,
    deadlock-on) using init ; parallelism{lock} ; [...] ; deadlock-on .

Solution 1
rewrites: 479677 in 2674887330ms cpu (7379ms real) (0 rewrites/second)
result Configuration: deadlock

No more solutions.
rewrites: 479677 in 2674887330ms cpu (7453ms real) (0 rewrites/second)
\end{verbatim}
}


\begin{thebibliography}{10}
\vspace{-5pt}

{
\bibitem{Caltais-aliasing}
G.~Caltais.
\newblock Expression-based aliasing for OO-languages.
\newblock Accepted in {\em 3rd International Workshop on Formal Techniques for Safety-Critical Systems 2014}; to appear.
\newblock {\em CoRR}, abs/1409.7509, 2014.

\bibitem{Coffman:1971:SD:356586.356588}
E.~G. Coffman, M.~Elphick, and A.~Shoshani.
\newblock System deadlocks.
\newblock {\em ACM Comput. Surv.}, 3(2):67--78, 1971.

\bibitem{DBLP:conf/ifm/GiachinoGLLW13}
E.~Giachino, C.~A. Grazia, C.~Laneve, M.~Lienhardt, and P.~Y.~H. Wong.
\newblock Deadlock analysis of concurrent objects: Theory and practice.
\newblock In {\em Integrated Formal Methods}, 394--411,
  2013.

\bibitem{Meyer-aliasing-13}
A.~Kogtenkov, B.~Meyer, and S.~Velder.
\newblock Alias and change calculi, applied to frame inference.
\newblock {\em CoRR}, abs/1307.3189, 2013.

\bibitem{Landi:1992:USA:161494.161501}
W.~Landi.
\newblock Undecidability of static analysis.
\newblock {\em ACM Lett. Program. Lang. Syst.}, 1(4):323--337,  1992.

\bibitem{Marti-OlietMV05}
N.~Mart\'{\i}-Oliet, J.~Meseguer, and A.~Verdejo.
\newblock {Towards a Strategy Language for Maude}.
\newblock In {\em Electr. Notes in Theor. Comp. Sci.}, 117:417--441,
  2005.

\bibitem{DBLP:conf/fct/MeseguerR11}
J.~Meseguer and G.~Rosu.
\newblock The rewriting logic semantics project: {A} progress report.
\newblock In {\em Fundamentals of Computation Theory},
  1--37, 2011.

\bibitem{DBLP:conf/acsd/MorandiSNM13}
B.~Morandi, M.~Schill, S.~Nanz, and B.~Meyer.
\newblock Prototyping a concurrency model.
\newblock In {\em 13th International Conference on Application of Concurrency
  to System Design}, 170--179, 2013.
}

\end{thebibliography}
\end{document}